\documentclass[aps,prd,twocolumn,groupedaddress,superscriptaddress,preprintnumbers,floatfix]{revtex4-1}
\usepackage{bm,pifont}
\usepackage{amssymb}
\usepackage{amsmath}
\usepackage{amsfonts}
\usepackage{epsfig}
\usepackage{graphicx}
\usepackage{tabularx}
\usepackage{color}
\newcommand{\seq}{\begin{subequations}}
\newcommand{\sen}{\end{subequations}}
\newcommand{\eq}{\begin{eqnarray}}
\newcommand{\en}{\end{eqnarray}}
\newcommand{\la}{\langle} 
\newcommand{\ra}{\rangle} 

\def\shiftdown#1{#1\llap{\lower.04ex\hbox{#1}}}

\begin{document}
\title{Deuteron EDM induced by CP violating 
couplings of pseudoscalar mesons}
	
\date{\today}
	
\author{Alexey S.~Zhevlakov} 
\email{zhevlakov@phys.tsu.ru}
\affiliation{
Department of Physics, Tomsk State University, 634050 Tomsk, Russia} 
\affiliation{Matrosov Institute for System Dynamics and 
Control Theory SB RAS Lermontov str., 134, 664033, Irkutsk, Russia } 
\author{Valery E. Lyubovitskij}
\email{valeri.lyubovitskij@uni-tuebingen.de} 
\affiliation{Institut f\"ur Theoretische Physik,
Universit\"at T\"ubingen,
Kepler Center for Astro and Particle Physics,
Auf der Morgenstelle 14, D-72076 T\"ubingen, Germany}
\affiliation{Departamento de F\'\i sica y Centro Cient\'\i fico
Tecnol\'ogico de Valpara\'\i so-CCTVal, Universidad T\'ecnica
Federico Santa Mar\'\i a, Casilla 110-V, Valpara\'\i so, Chile}
		
\begin{abstract}

We analyze contributions to the electric dipole (EDM) and Schiff (SFM) 
moments of deuteron induced by the CP-violating three-pseudoscalar meson 
couplings using phenomenological Lagrangian approach 
involving nucleons and pseudoscalar mesons $P=\pi, K, \eta, \eta'$. 
Deuteron is considered as a proton-neutron bound state and its properties 
are defined by one- and two-body forces. One-body forces correspond to 
a picture there proton and neutron are quasi free constituents of deuteron 
and their contribution to the deuteron EDM (dEDM) 
is simply the sum of proton and neutron EDMs. 
Two-body forces in deuteron are induced by one-meson 
exchange between nucleons. They produce a contribution to the dEDM, which 
is estimated using corresponding potential approach. 
From numerical analysis of nucleon and deuteron EDMs we derive stringent limits 
on CP-violating hadronic couplings and $\bar\theta$ parameter. 
We showed that proposed measurements of proton and deuteron  EDMs 
at level of $\sim 10^{-29}$ by the Store Ring EDM and JEDI Collaborations 
will provide more stringent upper limits on the CP violating parameters. 
  	
\end{abstract}
	
\maketitle
	
\section{Introduction}

Study of nature of CP-violation is one of the most important tasks in particle physics. 
Here the main puzzle consists in disagreement of predictions of Standard Model (SM) 
and existing data on CP-violating effects like, e.g., electric dipole moments (EDMs) 
of electron, nucleons, and  more composite system like deuteron and nuclei. 
SM gives more stringent upper limits than experiments. 
It calls for search for a New Physics (new particles or mechanisms) contributing 
to CP-violating effects. In particular, data bounds on the hadron and lepton EDMs 
are very useful for derivation of more stringent limits on parameters 
of new particles~\cite{Kirpichnikov:2020tcf,Kirpichnikov:2020new}. 
In QCD the source of the CP-violation is encoded in the so-called 
QCD vacuum angle $\bar\theta$, 
which is very small quantity ($\bar\theta \sim 10^{-10}$) due to Peccei-Quinn 
mechanism~\cite{Peccei:1977hh}. 
As it was shown in QCD sum rules~\cite{Crewther:1979pi,Shifman:1979if},  
this angle is related to the effective CP-violating hadronic couplings, which, 
e.g., define the EDMs of baryons. E.g., the expressions for the CP-violating 
$\eta(\eta')\pi\pi$ couplings derived 
in Refs.~\cite{Crewther:1979pi,Shifman:1979if} read: 
\eq\label{QCD_fetapipi}
f_{H\pi\pi} &=& - g_H \,
\frac{\bar\theta\, M_\pi^2\, R}{F_\pi\, M_H \, (1+R)^2}\,, \quad H=\eta, \eta' 
\,,\label{eq:f-theta}
\en
where $g_\eta = \sqrt{1/3}$, $g_{\eta'} = \sqrt{2/3}$, 
$R = m_u/m_d$ is the ratio of the $u$ and $d$ current quark masses,
$F_\pi=92.4$ MeV is the pion decay constant, $M_\pi=139.57$ MeV,  
$M_\eta = 547.862$ MeV, and $M_{\eta'} = 957.78$ MeV 
are the masses of the charged pion, $\eta$, and $\eta'$ mesons, 
respectively. 
   
In series of papers~\cite{Gutsche:2016jap}-\cite{Zhevlakov:2019ymi}
we developed phenomenological Lagrangian approach involving nucleons, pseudoscalar 
mesons $P = \pi, \eta, \eta'$, and photon for analysis of nucleon EDM and deriving 
upper limits for the CP-violating couplings between hadrons and $\bar\theta$ angle. 
In particular, using existing upper limit on neutron EDM (nEDM)~\cite{Tanabashi:2018oca} 
\begin{align}
|d_n| <  2.9 \times 10^{-26} \,\text{e} \cdot \text{cm}\,,
\end{align}	 
which corresponds to the following boundary for the QCD angle 
$|\bar\theta| < 10^{-10}$, we derived more stringent 
upper limits for the CP-violating $\eta\pi\pi$ and $\eta'\pi\pi$ 
couplings $f_{\eta\pi\pi} < 4.4 \times 10^{-11}$ and 
$f_{\eta'\pi\pi} < 3.8 \times 10^{-11}$ than the ones deduced from experiment 
by the LHCb Collaboration~\cite{Aaij:2016jaa}: 
$f_{\eta\pi\pi} < 6.7 \times 10^{-4}$ and $f_{\eta'\pi\pi} < 2.2 \times 10^{-4}$. 
Using limits for these coupling one can estimate other hadronic EDMs where 
these couplings contribute. The proposed experiments for measurement of 
EDMs of charge particles (proton, deuteron, and possibly helium-3) 
with a sensitivity of $10^{-29} e \cdot {\rm cm}$ by 
several Collaborations (the Storage Ring EDM at 
BNL~\cite{Anastassopoulos:2015ura}, the JEDI 
at J\"ulich\cite{Eversmann:2015jnk,Abusaif:2019gry}) call for 
more accurate theoretical analysis of EDMs. 

In this paper we extend our analysis to the deuteron, which is considered as proton-neutron 
bound state. In addition to the deuteron EDM (dEDM) 
we estimate the slope of the EDM form factor, which 
is known as the Schiff moment. We will take into account the contributions of 
one- and two-body forces to the dEDM. 
One-body forces correspond to a picture there proton and neutron are quasi free 
constituents of deuteron and their contribution to the dEDM  
is simply the sum of proton and neutron EDMs. 
Two-body forces in deuteron are induced by one-meson ($\pi$, $\eta$, and $\eta'$) 
exchange between nucleons. They produce a contribution to the dEDM, which 
is estimated using potential approach proposed in Ref.~\cite{Khriplovich:1999qr}.   
From numerical analysis of nucleon and deuteron EDMs we derive stringent limits 
on the CP-violating hadronic couplings and $\bar\theta$ parameter. 
We show that proposed measurements of proton and deuteron  EDMs 
at level of $\sim 10^{-29}$ by the Store Ring EDM and JEDI Collaborations 
will provide more stringent upper limits on the CP-violation parameters. 

The paper is organized as follows.
In Sec.~II we briefly discuss our formalism and 
results for EDM and Schiff moments of nucleons.
In Sec.~III we extend our formalism to the dEDM. 
In Sec.~IV we present our numerical results for the dEDM and 
discuss it in connection with planned experiments.
In Appendix~A we present the results for the $K$- mesons contributions to 
the pseudoscalar meson and baryon CP-violating couplings relevant for 
the $|\Delta T|=0,1$ isospin transition.

\section{Formalism} 
	
\begingroup
\squeezetable
\begin{figure}[t!]		
\includegraphics[width=0.45\textwidth, trim={3cm 19.8cm 12cm 2cm},clip]{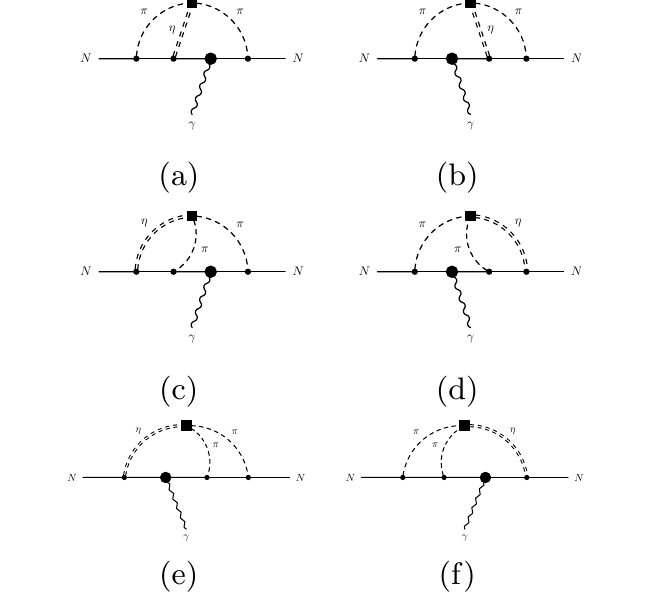}
	\caption{Diagrams describing the nEDM induced 
                 by the minimal electric coupling of photon with charged baryon.
		 Interaction between mesons and baryons is described 
                 in the framework 
                 of PS approach. The solid square denotes the CP-violating 
                $\eta\,\pi^+\pi^-$ and $\eta\,\pi^0\pi^0$ vertices.} 
\label{two_loop_1}
\end{figure}
\endgroup	
	
In this section we briefly review our formalism, which is based on phenomenological Lagrangians 
formulated in terms of nucleons $N=(p,n)$, pseudoscalar mesons [pions $\pi=(\pi^\pm,\pi^0)$ and 
etas $H=(\eta,\eta')$], and photon $A_\mu$ (see details in Ref.~\cite{Zhevlakov:2018rwo}). 
The full Lagrangian needed for the analysis of nucleon EDMs is conventionally divided on 
free ${\cal L}_{0}$ and interaction ${\cal L}_{\rm int}$ parts ~\cite{Zhevlakov:2018rwo}. 
In particular, the interaction part ${\cal L}_{\rm int}$ is given a sum of 
CP-even and CP-odd strong interactions terms ${\cal L}_{S}$ and ${\cal L}_{S}^{\rm CP}$ 
and electromagnetic terms describing coupling of charged pions and nucleons with photon. 
In case of nucleons we take into account non-minimal coupling with photon induced by 
anomalous magnetic moment $k_N$: 
\eq\label{Lint} 
{\cal L}_{\rm int} &=& {\cal L}_{\pi NN} + {\cal L}_{H NN} 
+ {\cal L}^{\rm CP}_{H \pi\pi} 
+ {\cal L}_{\gamma NN} + {\cal L}_{\gamma\pi\pi}\,, \nonumber\\[2mm]
{\cal L}_{\pi NN} &=& g_{\pi NN} \bar{N} i\gamma^5 \vec{\pi\,} \vec{\tau\,} N\,, \ \  
{\cal L}_{HNN} \ = \ g_{HNN} H \bar{N} i\gamma^5 N\,, \nonumber\\[2mm]
{\cal L}^{\rm CP}_{H\pi\pi} &=& f_{H\pi\pi} M_H H \vec{\pi\,}^2 \,,\nonumber
\en
\eq
{\cal L}_{\gamma NN} &=& e A_\mu N \Big(\gamma^\mu Q_N 
+ \frac{i \sigma^{\mu\nu} q_\nu}{2 M_N} k_N \Big) N\,, 
\\
{\cal L}_{\gamma\pi\pi} &=& e A_\mu \Big(\pi^- i\partial^\mu \pi^+ 
- \pi^+ i\partial^\mu \pi^-\Big) 
+ e^2 A_\mu A^\mu \pi^+ \pi^- \,, \nonumber
\en  
where $g_{\pi NN} = (g_A/F_\pi) \, m_N$, 
$g_{HNN}$, and $f_{H\pi\pi}$ are corresponding coupling constants, 
$\gamma^\mu$, $\gamma^5$ are the Dirac matrices, 
$\sigma^{\mu\nu} = \frac{i}{2}[\gamma^\mu,\gamma^\nu]$. 
Here $g_A = 1.275$ is the axial nucleon charge, 
For the constants $g_{\eta NN}$ and $g_{\eta^\prime NN}$ we use the values deduced 
from recent analysis of photoproduction on nucleons in Ref.~\cite{Tiator:2018heh}: 
$g_{\eta NN} = g_{\eta^\prime NN} =~0.9$.

\begingroup
\begin{figure}	[t!]	
\includegraphics[width=0.5\textwidth, trim={2.5cm 22.6cm 11cm 1.5cm},clip]{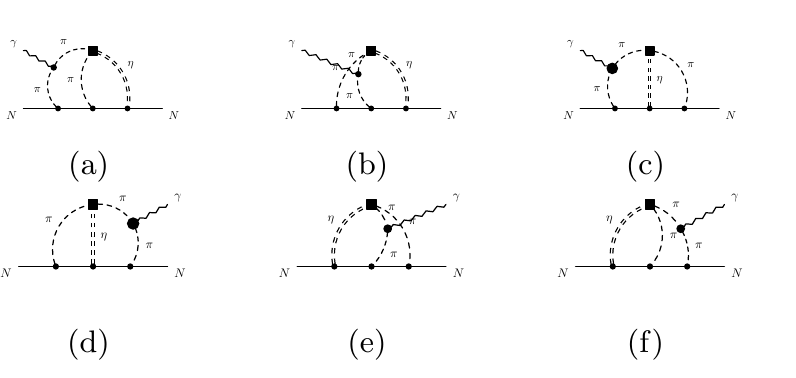}
	\caption{Diagrams describing the nEDM induced 
                 by the minimal electric coupling of photon with charged pions.
		 Interaction between mesons and baryons is described in the framework 
                 of PS approach. 
		 The solid square denotes the CP-violating $\eta\,\pi^+\pi^-$ vertex. }
	\label{two_loop_2}
\end{figure}
\endgroup	

Nucleon EDM is extracted from the electromagnetic vertex function, which is expanded 
in terms of four relativistic form factors $F_E(Q^2)$ (electric), 
$F_M(Q^2)$ (magnetic), $F_D(Q^2)$ (electric dipole), 
and $F_A(Q^2)$ (anapole) as~\cite{EDM_PCQM,EDM_SUSY}: 
\begin{align}
&M_{\rm inv}=\bar u_N(p_2)\, 
\Gamma^\mu(p_1,p_2)\,u_N(p_1)\,,
\\
&\Gamma^\mu(p_1,p_2) \,=\,  \nonumber
\gamma^{\mu} \, F_E(Q^2)  + \qquad\qquad\qquad \\ 
&
\,+\,\frac{i\sigma^{\mu\nu}}{2 m_N} q_{\nu} \, F_M(Q^2)\nonumber
+\frac{\sigma^{\mu\nu}}{2 m_N}q_{\nu} \gamma^5 \, F_D(Q^2) +\\&
\,+\, \frac{1}{m_N^2} (\gamma^\mu q^2 - 2m_N q^{\mu}) \gamma^5
\, F_A(Q^2) \, ,
\label{vertex}
\end{align}
where $p_1$ and $p_2$ are momenta of initial and final nucleon 
states, $Q^2=(p_2-p_1)^2$ is the transfer momentum squared. 
The nucleon EDM is defined as $d_N^E=-F_D(0)/(2 m_N)$. 

In preceding papers~\cite{Gutsche:2016jap}-\cite{Zhevlakov:2019ymi} 
we analyzed the neutron EDM (nEDM), 
which is evaluated by taking into account the two-loop 
diagrams. E.g., in Figs.~\ref{two_loop_1} and~\ref{two_loop_2} 
we display the diagrams induced by minimal couplings of charged hadrons 
with photon (see details in Ref.~\cite{Zhevlakov:2018rwo}). 
The contributions to the nEDM induced by non-minimal coupling of proton and neutron 
to the electromagnetic field have been analyzed in Ref.~\cite{Zhevlakov:2019ymi}. 
We showed that non-minimal contributions  
are of the same order of magnitude as the ones induced by minimal $\gamma$-proton coupling,  
but separate non-minimal contributions induced by anomalous magnetic moments of proton 
and neutron compensate each other due to their opposite sign.  
The total numerical contribution of the non-minimal 
couplings of the nucleon is relatively suppressed (by one order of magnitude) compared 
to the total 
contribution of the minimal coupling. Our final numerical results for the nEDM including 
minimal and non-minimal electromagnetic couplings of nucleons are~\cite{Zhevlakov:2019ymi}: 
\begin{align}
d^{E}_n\simeq(6.62 f_{\eta\pi\pi}
+7.64 f_{\eta^\prime\pi\pi})\times 10^{-16} \text{e} \cdot \text{cm} \, ,
\end{align} 
in terms of the CP-violating $\eta\pi\pi$ and $\eta'\pi\pi$ couplings and 
in terms of the QCD $\bar\theta$ angle: 
\begin{align}
|d^{E}_n| \simeq 0.64 \times 10^{-16} \bar\theta \,\text{e} \cdot \text{cm} \, ,
\end{align} 
using the 
ratio of $u$- and $d-$ quarks $R=0.556$ from ChPT at 1 GeV scale~\cite{Gasser:1982ap} 
and 
\begin{align}
|d^{E}_n| \simeq 0.67 \times 10^{-16} \bar\theta \,\text{e} \cdot \text{cm} \, ,
\end{align} 
for the $R=0.468$ taken from lattice QCD 
at scale of 2 GeV~\cite{Tanabashi:2018oca}. 
Then using data on the nEDM we deduced~\cite{Zhevlakov:2019ymi} 
the following upper limits on the QCD angle: 
$|\bar\theta|< 4.4 \times 10^{-10}$ (ChPT) and 
$|\bar\theta|< 4.7 \times 10^{-10}$ (lattice QCD). 

In this paper  we first do an extension of our formalism to the proton EDM (pEDM), 
which is straightforward. Our numerical results for the pEDM in terms 
of the CP-violating $\eta\pi\pi$ and $\eta'\pi\pi$ couplings are  
\begin{align}
d^{E}_p\simeq(1.66 f_{\eta\pi\pi}
+1.77 f_{\eta^\prime\pi\pi})\times 10^{-16} \text{e} \cdot \text{cm} \,. 
\end{align} 
Using relations of the $\eta\pi\pi$ and $\eta'\pi\pi$ couplings 
with $\bar\theta$ we express the pEDM in terms of the $\bar\theta$ as: 
\begin{align}
|d^{E}_p| \simeq 0.15 \times 10^{-16} \bar\theta \,\text{e} \cdot \text{cm} \, ,
\end{align} 
 for $R=0.556$ from ChPT  and same
\begin{align}
|d^{E}_p| \simeq 0.16 \times 10^{-16} \bar\theta \,\text{e} \cdot \text{cm} \, ,
\end{align} 
for $R=0.468$ from lattice QCD. The magnitude of pEDM is less then the one for nEDM 
because in case of pEDM the contributions from diagrams in Fig.~\ref{two_loop_1} 
and Fig.~\ref{two_loop_2} have different sign in comparison with their contribution 
to the nEDM. Due to the leading diagrams to the pEDM and nEDM are induced by 
the coupling of photon with charged pions with opposite charge, 
i.e. with $\pi^-$ and $\pi^+$, respectively, the nucleon EDMs should have 
different signs. 

After substitution of upper limits for the $\bar\theta$ derived in the neutron case 
we get the following upper limits for the pEDM: 
$|d^{E}_p| \simeq 0.72 \times 10^{-26} \,\text{e} \cdot \text{cm}$ (ChPT) and 
$|d^{E}_p| \simeq 0.68 \times 10^{-26} \,\text{e} \cdot \text{cm}$ (lattice QCD). 
These limits are more stringent than existing limit $|d^E_p|<2.5 \times 10^{-25}$  
obtained in indirect way from analysis of the Hg 
atoms~\cite{Griffith:2009zz,Sahoo:2016zvr,Dmitriev:2003sc} 
and have the same order of magnitude as the nEDM. 

We go further and estimate the Schiff moments (SFMs) of nucleons. 
The nucleon SFM is defined as the slope of its 
EDM form factor~\cite{Faessler:2005gd}:  
\begin{align}
S^\prime_N=-\frac{d_N^E(Q^2)}{dQ^2}\Bigg|_{Q^2=0} \, .
\end{align}
Our numerical results for the nucleon SFMs are: 
\begin{align}
&|S'_n|<(4.1 \, f_{\eta \pi\pi}+4.4 \, f_{\eta' \pi\pi})   \cdot 10^{-3}  
\, \text{e} \cdot \text{fm}^3 \, ,
\\
&|S'_p|< (3.7 \, f_{\eta \pi\pi}+3.9 \, f_{\eta' \pi\pi})   \cdot 10^{-3} 
\, \text{e} \cdot \text{fm}^3 
\end{align} 
in terms of the CP-violating $\eta\pi\pi$ and $\eta'\pi\pi$ couplings 
and in terms of the QCD vacuum angle 
\begin{align}
&|S'_n| < 3.9 \cdot 10^{-4} \, \bar\theta \, \text{e} \cdot \text{fm}^3 \, ,
\\
&|S'_p| < 3.6 \cdot 10^{-4} \, \bar\theta \, \text{e} \cdot \text{fm}^3 \, ,
\end{align}
for the ChPT set 
and  
\begin{align}
&|S'_n| < 3.7 \cdot 10^{-4} \, \bar\theta \, \text{e} \cdot \text{fm}^3 \, ,
\\
&|S'_p| < 3.4 \cdot 10^{-4} \, \bar\theta \, \text{e} \cdot \text{fm}^3 \, ,
\end{align}
for the lattice QCD set. Main contribution to the nucleon SFMs comes from 
the diagram describing the coupling of photon with charged pions 
(see Fig.~\ref{two_loop_2}), while the contribution of the graphs in 
Fig.~\ref{two_loop_1} is suppressed. It is different from the nucleon EDMs, 
which are generated by both sets of diagrams in Figs.~\ref{two_loop_1} 
and~\ref{two_loop_2} with equal contribution on magnitude. 

Note that our result for the neutron SFM is in good agreement 
with prediction of ChPT at the leading-order in 
the chiral expansion~\cite{Thomas:1994wi}:   
\eq
|S'_n| = \frac{e g_{\pi NN} g_{\pi NN}^{CP}}{48 \pi^2 M_\pi^2 m_N} 
< 4.4 \cdot 10^{-4} \, \bar\theta \, \text{e} \cdot \text{fm}^3 \, 
\en 
and perturbative chiral quark model~\cite{EDM_PCQM}:    
\eq  
|S'_n| < 3.0 \cdot 10^{-4} \, \bar\theta \, \text{e} \cdot \text{fm}^3 \,. 
\en 

\section{Deuteron EDM}

\begin{figure}[t]	
 \includegraphics[width=0.45\textwidth, trim={2cm 21cm 11cm 2cm},clip]{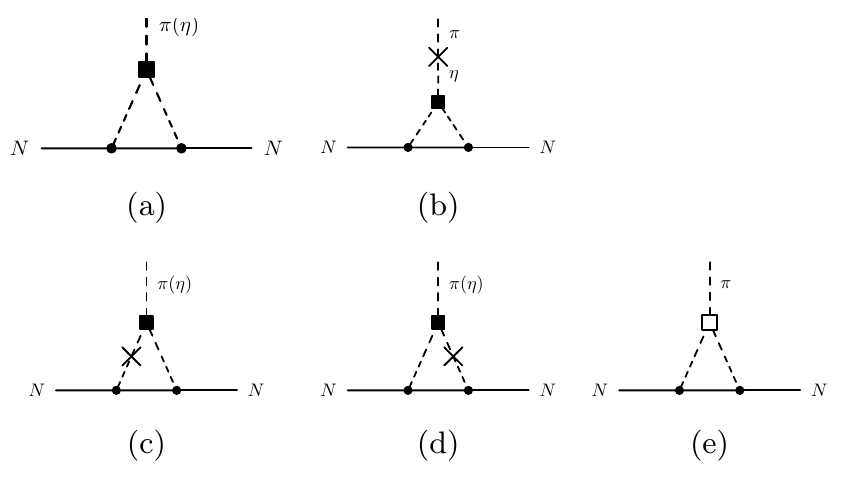}
\caption{CP-violating $\pi NN$ and $\eta NN$ couplings:  
(a) diagram induced by isospin-symmetric  ($|\Delta T| = 1$) vertices; 
(b) and (c) diagrams induced by isospin-violating ($|\Delta T| = 1$) 
vertices induced by the {\it internal} $\pi^0-\eta$ mixing; 
(d) diagram induced by isospin-violating ($|\Delta T| = 1$) 
pion-nucleon vertex and by the {\it external} $\pi^0-\eta$ mixing. 
The cross symbol $\times$ denotes the $\pi^0-\eta$ mixing; 
(e) diagram induced by CP-violating isospin-breaking coupling of 
three pseudoscalar mesons from Lagrangian~(\ref{CP-3m}). 
The black box symbol denotes the CP-violating isospin-symmetric $P^3$ vertex.  
The white box symbol denotes the CP-violating isospin-breaking $P^3$ vertex. 
\label{CPvert}}
\end{figure}

The CP-violating $HNN$, $H=\eta,\eta'$ couplings have been calculated in ChPT 
in Ref.~\cite{Gutsche:2016jap}.  
It is defined by pion-loop diagram and its value at the leading order 
in chiral expansion reads:  
\eq 
g_{HNN}^{\rm CP} 
&=& - \frac{3 g_A^2 f_{H\pi\pi}}{16 \pi^2 F_\pi^2} \, 
\, M_{H} \, m_N \nonumber\\
&=& - \frac{3\, g_{\pi NN}^2 \, f_{H\pi\pi}}{16 \pi^2} \, \frac{M_H}{m_N}\,.  
\en 
Here, by analogy we also calculate 
the CP-violating $\pi NN$ coupling, which is generated by similar loop 
diagram in Fig.~\ref{CPvert}a: 
\eq 
g_{\pi NN}^{\rm CP} &=& g_{\pi NN}^{\rm CP} (M_\eta) 
+ g_{\pi NN}^{\rm CP} (M_{\eta^\prime})\,,
\en
\eq
g_{\eta NN}^{\rm CP} = g_{\eta NN}^{\rm CP} (M_\pi)\,, \qquad 
g_{\eta^\prime NN}^{\rm CP} = g_{\eta^\prime NN}^{\rm CP} (M_\pi)\,,
\en
\eq 
g_{\pi NN}^{\rm CP} ( M_H) &=& 
- \frac{g_{\pi NN} g_{H NN} \, f_{H\pi\pi}}{4 \pi^2} \, \frac{M_H}{m_N} 
\nonumber\\
&\times&  
\biggl[ 1 + \frac{A(M_H^2) - A(M_\pi^2)}{2 m_N^2 (M_H^2 - M_\pi^2)}\biggr] 
\,,
\en\eq
g_{H NN}^{\rm CP} (M) &=& 
- 3 \, \frac{g_{\pi NN}^2 \, f_{H\pi\pi}}{16 \pi^2} \, \frac{M_H}{m_N} \, 
\biggl[ 1 + \frac{B(M^2)}{m_N^2} \biggr] 
\,,  \,\,\, \,\,\,
\label{gepapipi}
\en 
where 
\eq 
A(M^2) &=& M^4 \log\frac{m_N^2}{M^2} - M^3 \sqrt{4m_N^2 - M^2} \, C(M)\,, \nonumber \\
B(M^2) &=& M^2 \log\frac{m_N^2}{M^2} - 2 M \frac{3m_N^2 - M^2}{\sqrt{4m_N^2 - M^2}} 
\, C(M)\,, \nonumber\\
C(M) &=& \arctan\frac{2m_N^2-M^2}{\sqrt{4m_N^2M^2-M^4}} \nonumber\\
&+& \arctan\frac{M}{\sqrt{4m_N^2-M^2}} \,. 
\en 
 
As it is seen, that 
the CP-violating $\eta NN$ coupling dominates over the $\pi NN$ one. 
It is why it makes sense to take into account the $\eta$ exchange in the 
evaluation of the dEDM. Note that our CP-violating couplings have 
microscopic (loop) origin. Therefore, it is interesting to compare 
our loop results for the CP-violating meson-nucleon couplings 
with the results for these couplings derived using chiral techniques. 
In particular, the CP-violating $\pi NN$ coupling at the leading order 
in chiral expansion was obtained in Ref.~\cite{Crewther:1979pi} in terms 
of current quark masses and $m_\Xi - m_N$ baryon mass difference: 
\eq 
g_{\pi NN}^{\rm CP} = - \bar\theta 
\, \frac{(m_\Xi - m_N) \, R}{F_\pi \, (1 + R) \, (2 R_s - 1 - R)}\,,
\en 
where $R_s = m_s/m_d$ and $m_\Xi = 1321.71$ MeV is 
the mass of the $\Xi(1321)^-$ hyperon. 
Both $g_{\pi NN}^{\rm CP}$ and $g_{\eta NN}^{\rm CP}$ couplings 
can be presented in terms of matrix elements of quark operators projected 
over nucleon states (nucleon condensates)~\cite{Crewther:1979pi}: 
\eq 
g_{\pi NN}^{\rm CP} &=& - \bar\theta 
\, \frac{m_d R}{F_\pi \, (1 + R)} \, \la N|\bar q \tau^3 q |N\ra\, = -0.021 \, 
\bar\theta \,,  \,\,\,\\
g_{\eta NN}^{\rm CP} &=& - \frac{\bar\theta}{\sqrt{3}} 
\, \frac{m_d R}{F_\pi \, (1 + R)} \, \la N|\bar q q |N\ra\, =- 0.132 \,  
\bar\theta \, ,  \,\,\,\\
g_{\eta' NN}^{\rm CP} &=& - \bar\theta \sqrt{\frac{2}{3}} 
\, \frac{m_d R}{F_\pi \, (1 + R)} \, \la N|\bar q q |N\ra\,=- 0.28 \,  
\bar\theta  \, , \,\,\,
\,\,\,
\en 
for $R$ from lattice data at scale 2 GeV~\cite{Tanabashi:2018oca}. 
Matrix elements $\la N|\bar q \tau^3 q |N\ra$ and 
$\la N|\bar q q |N\ra$ can be related to the nucleon axial charge 
$g_A$~\cite{Faessler:2007pp} and pion-nucleon sigma-term 
$\sigma_{\pi N}$~\cite{Gasser:1990ce,Lyubovitskij:2000sf}: 
\eq 
& &\la N|\bar q \tau^3 q |N\ra = \frac{3}{5}\, g_A = 0.765 \,, \\
& &\la N|\bar q q |N\ra  = \frac{\sigma_{\pi N}}{\bar{m}} = 8.286\,, 
\en 
where $\bar{m} = (m_u + m_d)/2 = 7$ MeV~\cite{Gasser:1982ap}.  
For the $\sigma_{\pi N}$ we use 
the latest update 58 MeV derived in Ref.~\cite{RuizdeElvira:2017stg}. 

Our numerical results for the CP-violating constants in terms of 
the $\bar\theta$ parameter are: 
\eq
g_{\pi NN}^{\rm CP} &=& -0.021  \,  \bar\theta \, ,
\label{piNN_CP}\\
g_{\eta NN}^{\rm CP}   &=& -0.093  \,  \bar\theta \, ,
\label{etaNN_CP}\\
g_{\eta^\prime NN}^{\rm CP}   &=& -0.125  \,  \bar\theta \, .
\label{etapNN_CP} 
\en
including contribution of all pseudoscalar mesons 
in the loop diagrams (see details in Appendix~A) and 
\eq
g_{\pi NN}^{\rm CP} &=& -0.01  \,  \bar\theta \, ,
\\
g_{\eta NN}^{\rm CP}   &=& -0.11  \,  \bar\theta \, ,
\\
g_{\eta^\prime NN}^{\rm CP}   &=& -0.15  \,  \bar\theta \, ,
\en
when we restrict to the contribution of pion and $\eta$ meson 
in the loop diagrams. One can see, that our prediction for 
the full $g_{\pi NN}^{\rm CP}$ coupling~(\ref{piNN_CP}) is close 
to the prediction $0.027 \bar\theta$ of Ref.~\cite{Crewther:1979pi} 
and in agreement with central value of 
Ref.~\cite{Bsaisou:2012rg}: $(-0.018 \pm 0.007) \, \bar\theta$. 
On the other hand, the CP-violating constants 
$g_{\eta NN}^{\rm CP}$ and $g_{\eta' NN}^{\rm CP}$ dominate over 
$g_{\pi NN}^{\rm CP}$ by a one order of magnitude. 
They are so-called isospin $|\Delta T| = 0$ CP-violating 
couplings~\cite{Khriplovich:1999qr,Bsaisou:2012rg}. 

In this section we discuss calculation of the dEDM $d_D^E$, which 
is defined as the coupling 
of the external electric field $\vec{E}$ with deuteron spin 
$\vec{S}$: $H=-d_D^E \, (\vec{S}\cdot \vec{E})$. 
The contributions to the dEDM comes from one-body forces (additive 
sum of the pEDM 
and nEDM) and from two-body forces due one-meson exchanges between nucleons. 
The two-body contribution to the dEDM is induced by the CP-violating 
meson-nucleon coupling. Therefore, the dEDM is defined as: 
\begin{align}
d_D^E = d_p^E + d_n^E  + d_D^{\pi NN} ,
\end{align}
where $d_D^{\pi NN}$ is the two-body contribution due to 
pion meson exchange generated by $\pi^0-\eta$ and $\pi^0-\eta^\prime$ mixing. 

The two-body contributions can be estimated using potential approach 
proposed and developed in Ref.~\cite{Khriplovich:1999qr}. 
In case of the pion exchange it was shown that the dominant contribution 
comes due to the isospin triplet coupling~\cite{Khriplovich:1999qr}: 
\begin{align}
d_D^{\pi NN} = -\frac{e g_{\pi NN} \, 
g^{\rm CP(1)}_{\pi NN}}{12 \pi M_\pi} \, \rho_\pi\,, 
\label{chriplovich}
\end{align}
where 
\eq 
\rho_P= \frac{1+\xi_P}{(1+2\xi_P)^2}\,, 
\quad 
\xi_P= \frac{\sqrt{m_N \epsilon_D}}{M_P}\,. 
\en 
Here $P=\pi,\eta,\eta^\prime$ and $\epsilon_D = 2.23$ MeV 
is the deuteron binding energy, $g^{\rm CP(1)}_{\pi NN}$ is  
the CP-violating $\pi NN$ isospin-breaking coupling 
constant~\cite{Khriplovich:1999qr,Lebedev:2004va,Bsaisou:2012rg} 
including the $\eta-\pi$ and $\eta^\prime-\pi$ mixing 
\eq
g^{\rm CP(1)}_{\pi NN} = g^{\rm CP(\pi\eta)}_{\pi NN}+ 
g^{\rm CP(\pi\eta^\prime)}_{\pi NN} \,,
\en
\eq
g^{\rm CP(\pi\eta)}_{\pi NN} =\epsilon g^{\rm CP}_\eta(M_\eta)\,,  
\quad g^{\rm CP(\pi\eta^\prime)}_{\pi NN} =\epsilon^\prime 
g^{\rm CP}_{\eta^\prime} (M_\eta^\prime) \,, \,\,\,\,\,\,\, 
\label{gcp1}
\en 
where 
\eq
g^{\rm CP}_H(M_H) &=&  g^{\rm CP,Int}_H(M_H) + g^{\rm CP,Ext}_H(M_H)\, ,
\label{gcp2}
\en\eq
g^{\rm CP,Int}_H(M_H) &=&g^{\rm CP}_{\pi NN}-\frac{4}{3} 
g^{CP}_{H NN}(M_H)   \, , 
\label{gcp3}\\
g^{\rm CP,Ext}_H(M_H)&=& \biggl(\frac{\rho_H M_\pi}{\rho_\pi M_H} -1 \biggr)
\biggl(g_{H NN}^{CP}-g_{\pi NN}^{CP}\frac{g_{HNN}}{g_{\pi NN}}\biggr) 
 \,, 
\nonumber
\en
where $SU_f(3)$ flavor breaking coefficients $\epsilon$ and 
$\epsilon^\prime$~\cite{Leutwyler:1996np,Kroll:2004rs, Gardner:1998gz} 
defined as 
 \eq
 \epsilon=  \epsilon_0 \chi \cos\varphi \, ,
 \en
 \eq
\epsilon^\prime=  -2\epsilon_0 (1/\chi) \sin\varphi \, ,
 \en
with parameter $\epsilon_0$ encoding the isospin breaking effects:
\eq
 \epsilon_0= \frac{\sqrt{3}(1-R)}{2(2R_s-1-R)}\, ,
 \en
and $\chi=1+(4M_K^2-3M_\eta^2-M_\pi^2)/
(M_{\eta^\prime}^2-M_\eta^2) \simeq 1.23$. 
Here $\varphi \simeq -21.6^0$ 
is mixing  angle between $\eta$ and $\eta^\prime$ mesons which 
is fixed from relation $\sin2 \varphi = -(4\sqrt{2}/3)(M_K^2-M_\pi^2)/
(M_{\eta^\prime}^2-M_\eta^2)$~\cite{Leutwyler:1996np,Kroll:2004rs,%
Gardner:1998gz}. Resulting values are $\epsilon = 0.017$ and 
$\epsilon^\prime =0.004$.
 
First term in Eq.~(\ref{gcp2}) is induced by the $\pi^0-\eta$ mixing 
in the triangle loop diagrams in Figs.~\ref{CPvert}c and ~\ref{CPvert}d.  
Second term in Eq.~(\ref{gcp2}) is induced by $\pi^0-\eta$ mixing in the external 
meson leg (see Fig.~\ref{CPvert}b). Therefore, one can denote two 
mechanisms of isospin violation due to the $\pi^0-\eta$ mixing as 
{\it internal mechanism} (depicted in Figs.~\ref{CPvert}c 
and~\ref{CPvert}d) and as {\it external mechanism} 
(depicted in Fig.~\ref{CPvert}b). 
One should note that the {\it internal mechanism} is strongly suppressed 
in comparison with {\it external mechanism} by a factor $10^{-2}$ . 
We get the following numerical results for 
the CP-violating $g^{\rm CP(1)}_{\pi NN}$ coupling: 
\eq
g^{\rm CP(\pi\eta)}_{\pi NN} = 0.0016 \, \bar\theta 
\en
due to $\pi-\eta$ mixing 
and
\eq
g^{\rm CP(\pi\eta^\prime)}_{\pi NN} = 0.0005\,  \bar\theta 
\en
due to $\pi-\eta^\prime$ mixing without $K$-mesons contribution. 
The total result with taking into account $K$-mesons contribution 
(see details in Appendix A) for the isospin breaking $|\Delta T|=1$ 
CP-violating pion-nucleon coupling is 
$g^{\rm CP(1)}_{\pi NN} = 0.0025 \,\bar\theta$, which is 
in good agreement with prediction 
of Refs.~\cite{Bsaisou:2012rg,Dekens:2014jka}: 
$\bar{g}^1 = (0.003\pm0.002)\, \bar\theta$.  

The ratio of the full CP-violating $\pi NN$ coupling constants 
corresponding to the $|\Delta T| = 1$ and $|\Delta T| = 0$ is: 
\eq
R_{\pi NN} = \frac{g^{\rm CP(1)}_{\pi NN}}{g^{\rm CP}_{\pi NN}} 
= \frac{\bar{g}^1}{\bar{g}^0}= -0.12\,. 
\en 
The latter expression also gives the prediction for the ratio of 
the couplings $\bar{g}^1$ and $\bar{g}^0$. 
One can see that our result for the $R_{\pi NN}$ is close to 
the lower boundary of the prediction of Ref.~\cite{Bsaisou:2012rg}: 
$R_{\pi NN} \sim -0.2\pm0.1$. 

Finally, resulting contribution from one-meson exchange is: 
\eq
|d_D^{\pi NN}| &=& 0.28 \cdot 10^{-18} \,  \bar\theta  \, \text{e cm}\,, 
\en
which is in good agreement with data~(see Ref.~\cite{Bsaisou:2012rg}). 
 
\section{Discussion}

The dEDM is contributed by the EDMs of constituent nucleons and 
correction due to one-meson exchange in the isospin channel $|\Delta T| = 1$. 
The latter is induced due to isospin breaking effects ($\eta-\pi$ and 
$\eta^\prime-\pi$ mixing) and, therefore, it is relatively suppressed. 
Our final prediction for the dEDM in terms of the $\bar\theta$ angle 
reads: 
\begin{align}	
|d_D|= 0.482\cdot 10^{-16} \,  \bar\theta \, \text{e cm}\,.
\end{align}	
Next using the upper limit for the $\bar\theta$~\cite{Zhevlakov:2019ymi} 
we get 
\begin{align}	
|d_D| < 2.2\cdot 10^{-26}  \, \text{e cm}\, . 	
\end{align}
Here we take into account that proton and neutron EDMs have different signs. 
 
In prospects of future experiments an observation that the dEDM is 
proportional to the nucleon EDM and the other contributions are suppressed  
has big importance. A comparison between our theoretical prediction and 
sensitivity of future experimental measurements of the dEDM 
at the level of accuracy $\sim 10^{-29}$ from the EDM 
Collaboration~\cite{Anastassopoulos:2015ura} gives more stringent 
limit on the CP-violating parameter $\bar\theta$: 
\begin{align}	
| \bar\theta | <  2 \cdot 10^{-13} \,. 	
\end{align}	
The same order of magnitude for the $\bar\theta$ has been obtained 
in framework of supersymmetric approach MSSM~\cite{Lebedev:2004va}.  
These limits on the dEDM and $\bar\theta$ allow to derive new bounds on 
nucleon EDMs at level $10^{-29}$ and more stringent limits on the decay 
rates of the CP-violation processes $\eta \to \pi\pi$ and $\eta' \to \pi\pi$.  
In connection with planned EDM experiments one can derive the limits 
for the branching ratios of these rare processes decays at level 
$\sim 10^{-21}$ and $\sim 10^{-23}$ for $\eta$ and $\eta^\prime$ mesons, 
respectively. Direct observation of these decays at a such level of 
accuracy is impossible. There is the same situation in case of 
future experiment on measurement of the proton EDM by the JEDI 
Collaboration~\cite{Eversmann:2015jnk,Abusaif:2019gry}. We would like 
to stress that direct measurement of the decay rates of the CP-violation 
processes $\eta \to \pi\pi$ and $\eta' \to \pi\pi$ at level higher 
than limitations from data on EDMs could potentially signal about 
manifestation of New Physics. 

In conclusion, we derived limits on the proton EDM and 
nucleon SFMs from existing experimental data on neutron EDM.  
We calculated the dEDM with taking into account one- and two-body 
forces in deuteron. All these quantities were calculated using 
phenomenological Lagrangian approach involving the PS-coupling between 
baryons and pseudoscalar mesons and the CP-violating couplings $3P$ 
couplings of pseudoscalar mesons. Note that these couplings are 
proportional to the QCD CP-violating parameter $\bar\theta$ and, 
therefore, encode a source of the strong CP-violation in our formalism. 
Complementary we also derived the dependence of the dEDM on $\bar\theta$.  
In future, we plan to continue our study of the EDMs of baryons and nuclei 
induced by strong CP-violating effects, e.g., by taking into account 
of the CP-violation three-pseudoscalar meson vertices involving all 
nonet states ($\pi$, $K$, $\eta$, and $\eta'$) and all isospin transitions 
$|\Delta T|=0, 1, 2$. 

\begin{acknowledgments} 

The work of A.S.Zh. was funded by Russian Science Foundation 
grant (RSF 18-72-00046). The work of V.E.L. was funded
by ``Verbundprojekt 05A2017-CRESST-XENON: Direkte Suche nach Dunkler 
Materie mit XENON1T/nT und CRESST-III. Teilprojekt~1 
(F\"orderkennzeichen 05A17VTA)'', by ANID PIA/APOYO AFB180002 
and by FONDECYT (Chile) under Grant No. 1191103.

\end{acknowledgments}	
	
\appendix
\section{SU(3) baryon-meson Lagrangian and CP-violating constants}

The baryon-meson interaction Lagrangian involving nucleon, $\Lambda$, and 
$\Sigma$ states in the framework of $SU(3)$ scheme 
reads~\cite{Stoks:1996yj, deVries:2015una} 
\eq
{\cal L}_{BBM} &=& g_{\pi NN} \vec{\pi} \bar{N} i\gamma_5 \vec{\tau} N 
+ g_{\Lambda NK} \Big(\bar{N}i\gamma_5\Lambda K + {\rm H.c.}\Big)
\nonumber \\
&+& g_{\Sigma NK} 
\Big(\bar{N} i \gamma_5 \vec{\Sigma} \vec{\tau} K + {\rm H.c.}\Big) \,, 
\en
where the relations between meson-baryon couplings are 
\eq
g_{\Lambda NK} &=& -\frac{g_{\pi NN}}{\sqrt{3}}
\frac{m_\Lambda+m_N}{2m_N}(1+\alpha)\, ,\\
g_{\Sigma NK} &=& g_{\pi NN}\frac{m_\Sigma+m_N}{2m_N}(1-2\alpha) \,,
\en
\noindent 
and $\alpha=F/(F+D)$. We use the averaged values for $F = 0.47$ 
and $D = 0.8$~\cite{Gutsche:2016jap,deVries:2015una,Ledwig:2014rfa} 
fixed from data. 

Effective Lagrangian for three meson couplings inducing 
the CP-violating processes with taking into account of 
isospin breaking effects reads~\cite{Crewther:1979pi,deVries:2015una}:
\eq
{\cal L}_{CP}
 &=& - \bar\theta \, \frac{M_\pi^2}{6 F_\pi}
\frac{m_u m_d }{(m_u+m_d)^2} \mathrm{Tr}(P^3) \,,
\en
where $P=P^a \lambda^a$ is the matrix of pseudoscalar fields.  
In terms of physical states this Lagrangian takes the form: 
\eq
{\cal L}_{CP} &=& -\frac{M_\pi^2 }{6 F_\pi \bar{m}}m^* 
\bar\theta \sqrt{3} \, \biggl[ \eta\, \vec{\pi}^2 
+ \sqrt{3} \, K^\dagger \vec{\pi} \vec{\tau} K 
-\eta \, K^\dagger K \nonumber\\
&+& \phi \, \biggl(\pi^0 \, \vec{\pi}^2
- \pi^0 \, K^\dagger K 
- \sqrt{3} \, \eta  \, K^\dagger \tau^3 K\biggr)   
\biggr] \\
&+& {\cal O}(\phi^2) \,, \nonumber
\label{CP-3m}
\en
where $\phi$ is the $SU(3)$ breaking parameter 
\eq
\phi &=& \frac{\sqrt{3}\bar{m}\tilde{\epsilon}}{2(m_s-\bar{m})} \,, 
\quad \bar{m} = \frac{m_u+m_d}{2} \, ,
\\
m^* &=& \frac{m_um_dm_s}{m_s(m_u+m_d)+m_um_d} \, , \quad 
\tilde{\epsilon} = \frac{m_d-m_u}{m_d+m_u} \,. \,\,\,\, \nonumber
\en
This Lagrangian generates the CP-violating couplings involving 
$\pi$, $K$, $\eta$ mesons and corresponds to the 
change of the isospin $|\Delta T|=0$ and $|\Delta T|=1$. 
Below we present the contribution of $K$-mesons to 
the $g^{CP}_{\pi NN}$, $g^{CP}_{\eta NN}$, and $g^{CP}_{\eta^\prime NN}$ 
couplings: 
\eq
g^{CP,K}_{\pi NN} &=& g^{CP,\Lambda}_{\pi NN} + g^{CP,\Sigma}_{\pi NN} \,,
\\
g^{CP,K}_{\eta(\eta') NN} &=&  -3 \left(g^{CP,\Lambda}_{\pi NN} 
+ g^{CP,\Sigma}_{\pi NN}\right)  \frac{f_{KK\eta(\eta')}}{f_{KK\pi}} \,,
\nonumber\\
g^{CP,B}_{\pi NN} &=&- \frac{g_{B NK}^2 f_{KK\pi}  (2m_N-m_B)}{16\pi^2 m_N^2}  \\
&& \times
[1+A(m_B) -C(m_B)]\,, 
\nonumber
\en 
where
\eq
A(m_B)& =&  \frac{m_B^2-M_K^2}{m_N^2} 
\log \left( \frac{m_B^2}{M_K^2} \right) \,,
\\
C(m_B) &=& \frac{ (M_K^2 - m_B^2)^2 - 
m_N^2 (M_K^2 + m_B^2) }{m_N^2\zeta} \,,
 \nonumber\\
&\times&
\left[
\arctan\left((m_B^2-M_K^2-m_N^2)\zeta^{-1}\right)  \right. \nonumber
\\
&-& \left.
\arctan\left( (m_B^2-M_K^2+m_N^2)\zeta^{-1}\right)  \nonumber
\right] \,,
\\
\zeta &=& \sqrt{2 M_K^2 (m_B^2+m_N^2)-(m_B^2-m_N^2)^2-M_K^4} \,, \nonumber
\en
where $B=\Lambda, \Sigma$ denotes hyperons and $f_{P_1P_2P_3}$ 
is the CP-violating three-pseudoscalar meson transition couplings  
obtained from Eq.~(\ref{CP-3m}).

The $K$-meson contribution to the $|\Delta T|=1$ CP-violating coupling shown 
in Fig.~\ref{CPvert}e is 
\eq
g_{\pi NN}^{CP,K} &=& \phi \left(  \frac{2}{3} 
\frac{f_{\pi^0\pi^+\pi^-}}{f_{\eta\pi\pi}} g^{CP}_{HNN}(M_\pi)
+   3g^{CP,K}_{\pi NN} \right) \,. \,\,\,\,\,\,\,\,\,
\en 
This contribution has the same magnitude as the value generated by the 
{\it internal mechanism} from diagrams Figs.\ref{CPvert}.c-\ref{CPvert}.d 
due to $f_{\eta\pi\pi}$ CP-violating coupling,  
the $ g^{CP}_{HNN}(M_\pi)$ function was denote before in eq.(\ref{gepapipi}).
Main contribution to $|\Delta T|=1$ CP-violated coupling of pion and nucleons due 
to $K$-mesons propagating in the loop also comes from the {\it external mechanism} 
(see Fig.~\ref{CPvert}b) which is given by the same structure integral: 
\eq
g^{\rm CP(1)}_{\pi NN} &=& g^{\rm CP,K(\pi\eta)}_{\pi NN} 
+ g^{\rm CP,K(\pi\eta^\prime)}_{\pi NN} \,, \\
g^{\rm CP,K (\pi\eta)}_{\pi NN} &=& 
\epsilon g^{\rm CP}_\eta(M_\eta)\,, \nonumber
\quad g^{\rm CP, K(\pi\eta^\prime)}_{\pi NN} =
\epsilon^\prime g^{\rm CP}_{\eta^\prime} (M_{\eta^\prime}) \,, 
\nonumber\\
g^{\rm CP,Ext}_H(M_H)
&=&\biggl(\frac{\rho_H M_\pi}{\rho_\pi M_H} - 1 \biggr)
\biggl(g_{H NN}^{CP,K} 
-g_{\pi NN}^{CP,K}\frac{g_{HNN}}{g_{\pi NN}}\biggr)  \,, 
\nonumber
\en
where $H = \eta, \eta^\prime$. It contributes by amount of $15\%$ 
to the CP-violating $\pi NN$ coupling in case of the $|\Delta T|=1$ 
isospin transition.

\end{document}